# INFORMATION ANALYSIS OF DNA SEQUENCES

Mohammed Riyazuddin

## Abstract

The problem of differentiating the informational content of coding (exons) and non-coding (introns) regions of a DNA sequence is one of the central problems of genomics. The introns are estimated to be nearly 95% of the DNA and since they do not seem to participate in the process of transcription of amino-acids, they have been termed "junk DNA." Although it is believed that the non-coding regions in genomes have no role in cell growth and evolution, demonstration that these regions carry useful information would tend to falsify this belief. In this paper, we consider entropy as a measure of information by modifying the entropy expression to take into account the varying length of these sequences. Exons are usually much shorter in length than introns; therefore the comparison of the entropy values needs to be normalized. A length correction strategy was employed using randomly generated nucleic base strings built out of the alphabet $\{A, T, G, C\}$ of the same size as the exons under question. Our analysis shows that introns carry nearly as much of information as exons, disproving the notion that they do not carry any information. The entropy findings of this paper are likely to be of use in further study of other challenging works like the analysis of symmetry models of the genetic code.

## Motivation

The complexity and information carrying capacity of DNA data makes genomic sequence analysis an attractive research area today. More than 90% of the genome is known to be non-coding DNA (introns) and only 3-5 % of the sequence is the coding region (exons). Richard Roberts and Phillip Sharp won the 1993 Nobel Prize in Physiology and Medicine for their discovery of introns. Although it is believed that the non-coding regions in genomes have no role in cell growth and evolution, demonstration that these regions carry useful information would tend to falsify this belief.

DNA sequence analysis presents challenges in applying finite sequence theory and provides opportunity to explore for improvement on existing techniques. Intron sequences have been regarded by some researchers as once active genes that were involved in the evolution process but do not have any useful function now, much like the vestigial organs in the human body that are remains of our evolutional history [11].

Increasing availability of DNA data on the internet makes it possible to implement statistical and other techniques on sequences of different organisms. The advancement in technology over the past decade or so has created greater interest in studying genes, cell replication and the complexity of DNA.

## Outline

This paper discusses and presents tools in information theory to study the structure of exons and introns and to allow a reasonable comparison. Shannon's entropy of a finite sequence was primarily used as an analysis tool and to implement a benchmarking technique.

Other techniques explored include Autocorrelation and Kak's randomness test. MATLAB and C programming tools were used in the implementation of these novel methods.

The DNA character strings need to be converted to numerical values to apply mathematical tools on them. The bases were numericalized by substitution for performing the autocorrelation tests. It has been suggested in previous work that the coefficients of the Walsh Transform may be used to study the degree of randomness of a sequence [14]. This technique, called as Kak's randomness test was implemented for the total coding region in a genome and an intron sequence of a similar length.

The Shannon's entropy of exon and intron sequences was calculated by breaking up the sequence into sub-sequences of length $L$. An entropy plot is obtained by varying $L$ and calculating the probabilities of the sub-strings each time. The entropy thus obtained is termed as the "block entropy" which we shall denote as $H(E_i)$. The entropy per character (base) of the sequence is then obtained by normalizing the block entropy with the respective $L$ i.e.

For $L = 3, H(E_3)$ is normalized as,

$$H(E_3)' = H(E_3)/3 \qquad (1)$$

The entropy plots of various intron and exon sequences show that the entropy converges on increasing the search strings length. It follows that the entropy of sequences of a fixed length is a function of the finiteness of the sequence. There is hence a need to normalize the entropy in order to make a generic comparison of entropy patterns for sequences of different lengths. For this purpose, we have used randomly generated sequences from the alphabet $\{A, T, G, C\}$ of length equal to the DNA sequence under analysis, to obtain a proportionate correction factor for benchmarking the entropy values. An ensemble of random sequences having the same length was used to obtain optimum values of the benchmarking entropy values.

**Introduction**

**About DNA**

The DNA (Deoxyribonucleic acid) molecule residing in the cell nucleus encodes information conventionally represented as a symbolic string over the alphabet $\{A, T, G, C\}$[2]. The DNA molecule has a complex double helical structure (figure 1) which is formed as a result of folding between single strands of DNA. The combination between single strands of DNA takes place according to "Watson-Crick complementarity" that says that the only permissible combinations between bases are A-T or T-A and C-G or G-C hence one strand can easily be used to predict the other in a double stranded chain.

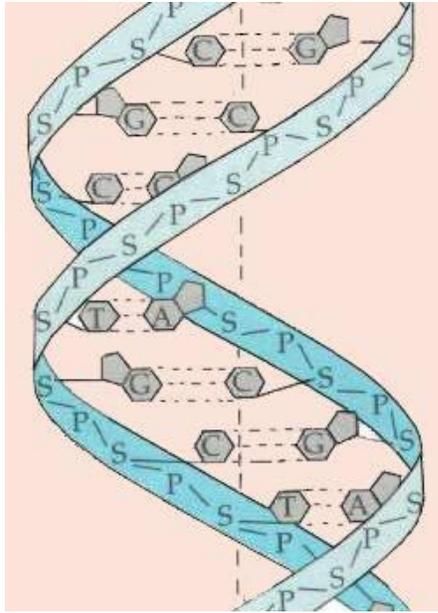

**Figure 1.** Double Helical structure of DNA

**Central dogma of a cell and genetic code**

The process of conversion of DNA to proteins involves the key stages: Transcription and Translation, according to Crick's Dogma of cell biology – figure 2.

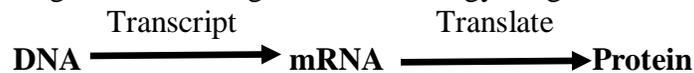

**Figure 2.** Central Dogma of Cell Biology (Crick)

The conversion of DNA to proteins may be visualized as a series of sequences generated in the process as shown in figure 3.

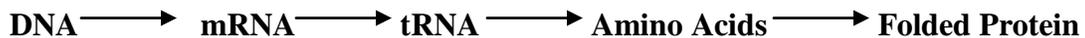

**Figure 3.** Genetic translation steps

The fact that large portions are removed during translation is regarded as one of the most unexpected findings in molecular biology [cited in 3]. During the translation of DNA to proteins, one or more of the codons map to one of the 20 amino acids according to the "Universal Genetic code" of the organism, resulting in a sequence of amino acids as shown in the example below.

*DNA Alphabet: {A,T,G,C}*
*DNA Sequence:* **ATGCCGCCCAAAACCCCCCGAA**…........
*Translated Protein Sequence:* **MPPKTPR**...

The length of a DNA sequence expressed in terms of base pairs (bp) varies from few thousands to several million bp. Although information in DNA sequence is normally analyzed using classical information theory, some quantum approaches have also been presented to better account for the structure. For quantum information and the logic behind its use in biological systems, see [5,6,7]; it has been suggested that codon symmetries can be better captured using quantum approach [8].

**Entropy**

Information Entropy was first introduced by Shannon [9]. Suppose $X$ be a random variable that assumes the values $x \in X$, $X$ being a finite set and the probability that $X$ assumes the particular value $x$ is denoted by $\Pr(x)$. Then the *Shannon entropy* of the random variable $X$ is defined as,

$$H(X) = -\sum_{x \in X} \Pr(x) \log_2 \Pr(x) \qquad (2)$$

The entropy $H(X)$ measures the average uncertainty in terms of bits of the outcome of the random variable $X$ [10]. The Shannon entropy is a measure of the order and disorder in sequences [4]. The entropy of a finite character sequence of length $N$ is defined as,

$$H(X) = \sum_i p_i \log(1/p_i) \qquad (3)$$

Where $i$ extends over all symbols of the alphabet,

$p_i$ is the probability of occurrence of symbol $s_i$ at any position.

And $p_i \in [0,1]$ for all $i = 1...N$ and $p_1 + ...p_N = 1$

For a given context, entropy is a measure of the order or disorder in a sequence that can be regarded as information [11]. The estimate of sequence entropy depends on the probabilities of words in the sequence and a general form of the probabilities is written as,

$$P(A) = \frac{n_A + \beta}{N + \beta d} \qquad (4)$$

Where $n_A$ is the frequency of event $A$ among $N$ total samples,

$d$ is the cardinality of the alphabet,

$\beta$ is a constant chosen as per case, $\beta = 0$ for normal maximum likelihood estimation and $\beta = 1$ was proposed by Laplace.

We shall now consider a chain of random variables $S_1, S_2, S_3...$ that range over a finite set $A$. This chain may be viewed as 1-D spin system, a stationary time series of measurements, or an orbit of a symbolic dynamical system [10]. The Shannon entropy for this block $S^L$ of variables may be defined as,

$$H(L) = -\sum_{s \in A} P(S_L) \log(S_L) \qquad (5)$$

**Importance of Entropy Estimation**

The Shannon Entropy of a data sequence is used to describe the complexity, compressibility, amount of information, weight of noise component, and so on. The DNA character strings are formed of the 4-letter alphabet $\{A, T, G, C\}$. Techniques have previously been used to apply the information theoretic notion of entropy to estimate the entropy of DNA sequences. It makes intuitive sense that the entropy of exons and introns differ since they are subject to different random processes [9]. Based on novel entropy estimation methods, issues like intron/exon boundary problem, the entropic difference of exons and introns, and the structure and information content of these sequences may be addressed. It has been demonstrated with tests on various genetic sequences that a significant difference exists between intron and exon entropies obtained using a match length entropy estimator [9]. This fast converging estimator was used to address the exon/intron boundary (splicing) problem extending the concept of indicators that represent the start and end of exon sequences. It was proved that a meaningful signal may be extracted from portions of a DNA sequence using this estimator. Another key result of the estimator was that the entropy of the gene sequences that actually code for proteins is higher compared to other DNA segments. This is in contrast to the biological theory prevalent at that time which explained that introns are capable of tolerating random sequences to a higher degree than exons.

**Entropy Estimation Techniques**

There are several methods for estimating the entropy of a random process. The most straightforward would be to find a direct computation of the expected log of the empirical distribution function. An entropy estimate thus obtained might only be as accurate as the estimate of the probability of n-tuples where n may be large. The entropy estimation is made difficult because of the shortness of the DNA sequences that code for proteins since the amount of data is practically insufficient to achieve a good estimate of all but the marginal or first order distribution and perhaps the distribution of pairs [9]. Data compression techniques like Lempel-Ziv (LZ) algorithm is another popular choice for entropy estimation. It is however known to have a slow rate of convergence for this purpose. Most of the techniques involve string matching and pattern frequency as part of the calculation. A match length entropy estimator has been proved to have a fast convergence rate relative to other techniques [9].

**CORRELATION AND RANDOMNESS TESTS**

Knowing and understanding the correlation between bases appearing in DNA sequences has been an interest for a long time now [16]. We begin with an introduction to the framework of correlation of random processes and later apply the same to DNA sequences.

**Mathematical Framework for Correlation**

**Autocorrelation Function:** Autocorrelation is the expected value of the product of a random variable or signal realization with a time-shifted version of itself. Assuming that we have two instances of the same random variable $X$ as $X_1 = X(t_1)$ and $X_2 = X(t_2)$, the autocorrelation of $X$ is written as,

$$R_{XX}(t_1 t_2) = E[X_1 X_2] = \int_{-\infty}^{\infty} \int_{-\infty}^{\infty} x_1 x_2 f(x_1, x_2) dx_2 dx_1 \qquad (6)$$

We will be applying autocorrelation to a real DNA sequence and we would like to look at the discrete time case of autocorrelation.

$$R_{XX}(n, n+m) = \sum_{-\infty}^{\infty} x(n)x(n+m) \qquad (7)$$

## DNA Sequence as a Random Process

Under the above framework, we will introduce the parameters involved in terms of DNA character sequences. Lets assume the DNA character sequence under question i.e. the exon or intron sequence to be a sample space 'S' of randomly occurring character strings and let the occurrence of a base at a given position k in the genome be a discrete random variable 'X'. We shall attempt to illustrate that there is underlying structure and patterns in DNA sequences. A simple proof of structure is the unequal probabilities of the characters $\{A, T, G, C\}$ obtained for sample exon and intron sequences taken from HUMRETBLAS are as follows:

For an Exon sequence of length 137 bases,
P(A) = 0.18, P(T) 0.06, P(C) = 0.47 and P(G) = 0.29
For an Intron sequence of length 3227 bases,
P(A) = 0.29, P(T) = 0.30, P(C) = 0.19 and P(G) = 0.23

The calculation and results of correlation for real DNA sequences are presented in the next section.

## Autocorrelation Plots of DNA Sequences

The Autocorrelation of a sequence is defined using Equation 3.5 as,

$$R_{XX}(n, n+m) = \sum_{-\infty}^{\infty} x(n)x(n+m)$$

Where $x(n)$ is discrete sequence of length.

The DNA character sequence under analysis has to be converted to a numeric sequence to apply mathematical functions on it. We are primarily interested in the autocorrelation of the sequences. Since autocorrelation is related to the pattern of letters in the sequence, we can use substitution to generate a numeric sequence. The characters {A,T,G,C} in the given sequence are substituted with arbitrary numerical values {-.5,.5,-1.5,1.5}. The substitution is done suitably to satisfy Watson-Crick's property which says A is complimentary with T and C with G. Numerical values are chosen such that correlation plots are symmetric about zero and hence easy to visualize.

A normalized form of the autocorrelation function was implemented in MATLAB by fixing the number of lags m to 10. For a discrete sequence of length N, the function used is represented mathematically as,

$$R_{xy}(m) = \frac{1}{N} * R_{xy}(m) \qquad (8)$$

where m is the number of lags and N the length of the sequence under test.

**Homosapiens genome:** (Genome Length = 16569 bp)

        Gene sequence (L = 957 bp)          Random sequence (L = 957)

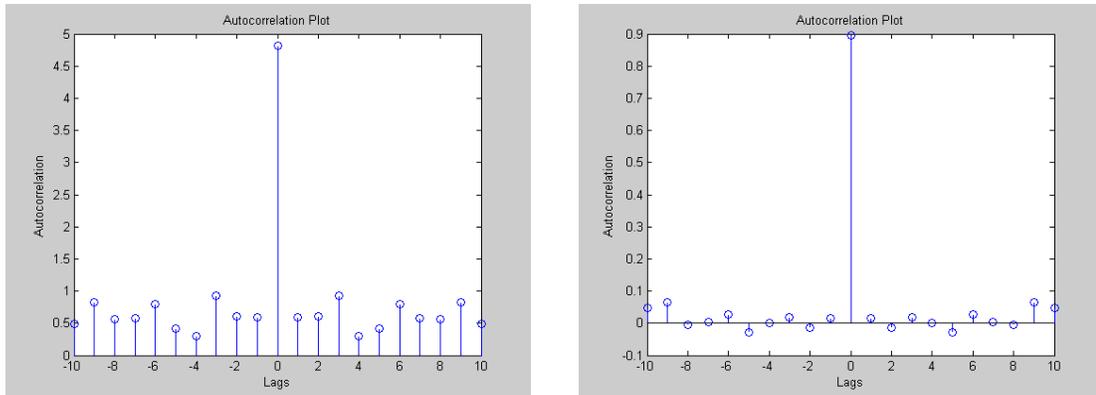

**Figure 4.** Autocorrelation of genetic sequence compared to a random sequence

It is evident from the above figure that a gene sequence shows structure compared to a random sequence of an equal length. The plot obtained is symmetric about the zero lag due to the symmetrical numerical values used in the sequence.

Exons and introns sometimes form part or whole of a gene sequence. We would be interested in studying the pattern and structure of the coding (exon) and non-coding regions (intron) in a DNA sequence. A comparison between their autocorrelation plots was our first test in this direction. Figure 5 shows autocorrelation plots for exon and intron sequences of similar length chosen from the HUMRETBLAS genome. A randomly generated sequence of the same length as the DNA sequences is used for comparison.

**Humretblas genome:** (Genome length = 180,388 bp)

        Exon (L = 77 bp)                        Intron (L = 77 bp)

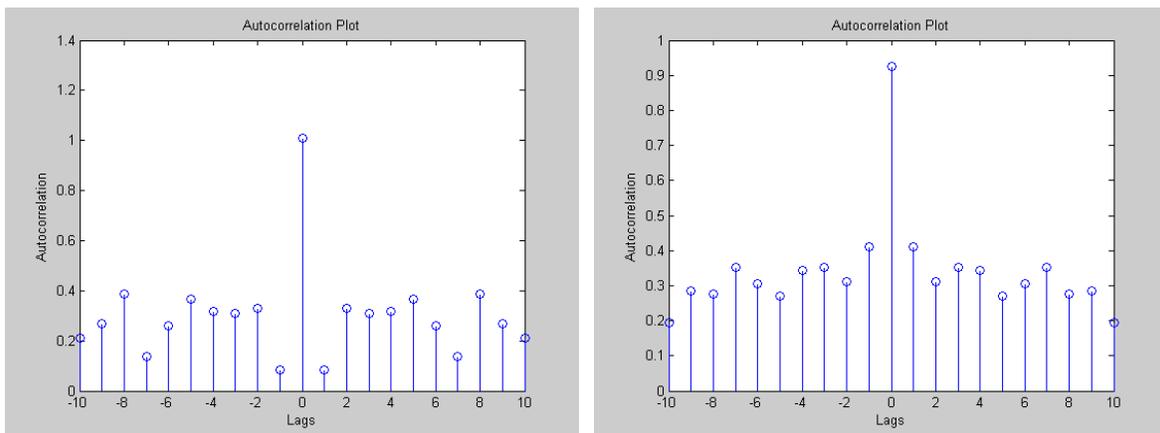

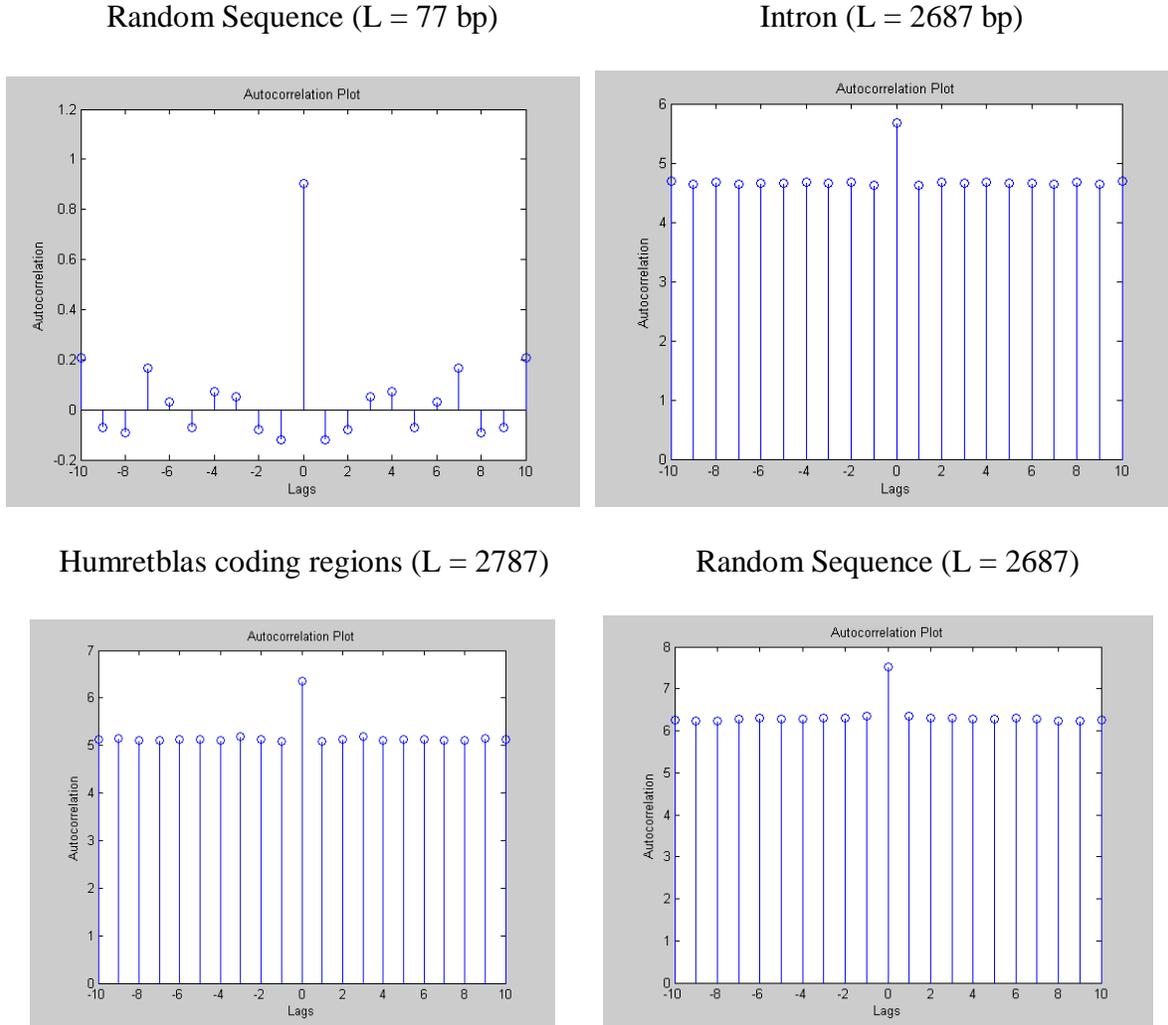

**Figure 5.** Autocorrelation plots of Introns and Exons

The amplitudes of autocorrelation at each lag point for exon and intron sequences of equal length were observed to differ only slightly and the pattern of the sequences seems identical. Intron sequences seem to carry meaningful patterns perhaps carrying information useful to the cell. The autocorrelation plots of introns and exons of similar length also suggest they may have structure comparable to exon sequences that encode critical information which is used in the translation of DNA to proteins.

**Kak's Randomness Test**

According to [14], "A sequence shall be said to have no pattern or be random if the number of independent amplitudes in the Walsh-Fourier transform is equal to the length of the sequence itself, i.e., $2^k$." The measure of randomness r(s) shall therefore be defined in terms of Walsh Transform values in the frequency domain for the sequence under test i.e.,

$$r(s) = \frac{i(s)}{L(s)} \qquad (9)$$

where i(s) = no. of independent amplitudes W(s)

and L(S) =length of the sequence

In this, "The number of independent amplitudes of W(s) shall equal the number of its component Walsh waves."[14]

The Walsh amplitudes are calculated using the MATLAB function for Walsh Transform. Consider a sequence [1 2 1 1]. A sample output of the function is shown below:

*walsh1D([1 2 1 1]) = 1.2500    0.2500    -0.2500    -0.2500*

A simple C script was then used to count the number of independent amplitudes in this result.

Randomness Measure, r(s) = W(s)/L(s) = 3/4 = 0.75 [considering 0.25 and -0.25 as independent]

At this end, let us assume that the sequence is zero padded towards the end [ 1 2 1 1 0 0 0 0] to increase the length of the final sequence to 8 characters i.e. the Walsh function is now,

*walsh1D([1 2 1 1 0 0 0 0]) = 0.625 0.625 0.125 0.125  -0.125  -0.125  -0.125  -0.125*

Randomness Measure, r(s) = W(s)/L(s) = 3/4 = 0.75 [considering 0.25 and -0.25 as different]

This observation clearly indicates that zero padding a sequence does not change the outcome of this randomness measure. This technique can be used to make the size of the DNA sequence equivalent to a $2^k$.

### Randomness Test Results

The Walsh function was applied to chosen intron and exon sequences and random sequences of the same length. The length of the sequence was adjusted to the nearest $2^k$ value either by truncating or zero padding the sequence. For example, the exon sequence of length 197 bp was zero padded to increase its length to 256. The following table shows results of this test.

| Sequence | Actual Length in bp | Adjusted to nearest $2^k$ | Kak's randomness coefficient R(s)/W(s) |
|---|---|---|---|
| Humretblas | 1,80,388 bp | | |
| Exon 1 Random | 197 | 256 | 41/256 45/256 |
| Total Coding Region Random | 2787 | 2048 | 146/2048 150/2048 |
| Intron 1 Random | 2687 | 2048 | 137/2048 133/2048 |

**Figure 6.** Results of Kak's Randomness Test

Due to the short length of exons, it's difficult to make a proper comparison of their structure to that of introns. For this purpose, we have used the total coding region that was generated by combining all exons from Humretblas and has a length comparable to the introns. The total coding region and the non-coding (intron) sequence under analysis were truncated to the nearest value 2048 in order to apply the Walsh function. The results obtained indicate similar structure for the coding and non-coding region.

## BENCHMARKING FINITE SEQUENCE ENTROPY

### Background of Finite Sequence Entropy

Entropy can be used to calculate the degree to which finite sequences can be compressed without any loss of information or to study structure of finite sequences. Although the existence of correlations in a sequence reduces the uncertainty of the symbols yet to be observed, it's important to locate them and account for them in our estimation. The most straightforward method of estimation would be based on the frequency of block strings up to a certain length and estimating their probabilities according to it.

$$\hat{p}(s_1, s_2, ..., s_n) = \frac{n_{s_1, s_2, ..., s_n}}{N} \tag{10}$$

where $n_{s1, s2...sn}$ is the no. of occurrences of the word $s_1, s_2, ..., s_n$

Then the entropy estimator may be written as,

$$\hat{h} = \lim_{n \to \infty} \hat{H}_n / n \tag{11}$$

Shannon's theory is based entirely on probabilistic concepts and deals with average code lengths but has a drawback that it doesn't take into account the information needed to describe the probability distribution itself.

### Algorithm Used for Entropy Calculation

As we know from Shannon's Entropy, the mathematical formulation of entropy is,

$$H(X) = \sum_i p(x_i) \log p(x_i) \tag{12}$$

In the case of DNA character sequences comprised of letters from the 4-letter alphabet $\{A, T, G, C\}$, the entity $X$ is represented as $S$ and assumed to be either a coding gene sequence or a non-coding sequence which we will refer interchangeably with exons and introns respectively. In order to compute the entropy using the Shannon's entropy shown in (12), the given sequence $X$ of length $N$ can be assumed to be comprised of a set of sub-strings of equal length $L$. Let us denote each of the sub-strings by $g_i$ and the number of sub-strings that make the sequence of length $N$ be $n$. In general, the number n is equal to $N/L$. Then the probability of each such sub-string $p(g_i)$ will be computed directly based on the frequency of occurrence of the string within the given DNA sequence i.e.

$$p(g_i) = \frac{n(g)}{N} \text{ Or } \frac{n(g)L}{N} \tag{13}$$

The entropy of a genetic sequence can then be represented as,

$$H(S) = -\sum_{i=1}^{N/L} p(g_i) \log p(g_i) \tag{14}$$

A C script was used to parse DNA character sequences and to calculate the frequencies of occurrence $n(g)$ for all sub-strings $g_i$. The script takes as input the genetic code and DNA sequence under question and implements a search algorithm to find the frequencies of occurrence of all the 64 possible codons from the genetic code in the input DNA sequence. The value of $L$ is then varied from 3 to 9 and a range of values of $H(X)$ is obtained which are then used to illustrate the behavior of the sequence entropy with increase in search lengths. For every $L$, the genetic code table is updated to include all the possible $L$-tuples $4^L$ comprised of characters from the alphabet $\{A,T,G,C\}$ since the number of search strings increases with increase in $L$. The size of possible sub-string lengths in fact grows exponentially as follows:

No. of possible triplets: $4^3 = 64$,

No. of 4-tuples: $4^4 = 256$,

No. of 5-tuples: $4^5 = 1024$, and so on.

The entropy value obtained for different sub-string lengths is sometimes called as the block entropy and is mathematically represented by:

$$H(L) = -\sum_{s \in A} P(S_L) \log(S_L) \tag{15}$$

This value is normalized with the block length to calculate the entropy per base i.e.,

$$H(S) = \frac{H(L)}{L} \tag{16}$$

As mentioned earlier the convergence of the entropy estimator must be fast enough to accommodate the shortness of certain DNA sequences [9]. The number of possible substrings increases with increase in triplets from $\{A,T,G,C\}$ are $4^3 = 64$, the number of 4-tuples is $4^4 = 256$, the number of 5-tuples is $4^5 = 1024$, and so on. However, the length of sequence available is limited and most exon sequences are only about 200 bp long. If the DNA sequence were periodic to repeat itself, it would need a sequence length comparable to the set of possible sub-strings.

We have used the Humretblas genome that has a total length of 180388 bp from NCBI. The annotations provided by the database to indicate coding and non-coding regions were also used to obtain several exon and intron sequences for our analysis. Figure 7 below shows a sample of DNA data with the integers showing the position of bases in the sequences. The orientation or direction in which the sequence is read is indicated in the database by 'complement' and that needs to be observed before using the sequence.

The entropy results obtained are shown a tabular as well as a graphical form below. Figure 8 shows a tabular result of entropy values where as Figure 9 illustrates exon entropies on a graph where y-axis indicates the entropy and sub-string lengths L = 3, 4,…9 are on the x-axis. Entropy convergence is rapid for exons when compared to some of the introns. As seen from the values, perhaps there is much similarity in the entropy convergence pattern of the coding sequence with that of an intron of a similar length.

```
ORIGIN
    1 gtaagtagtt cacagaatgt tatttttcac ttaaaaaaaa agatttttat ggaataatct
   61 caaacatctt gatagttagg gttagtttga tcgattatag caggctactt cataaattaa
  121 gcccatagat ttaagtcctg tgtagattat ttatcttctc acaaagaaaa tagtataaaa
  181 tacatgcctt gtactacaaa gaagaactaa taaggtggaa ttgattcagg acagcatatc
  241 accaactctg agaaaaatgc aacaaatgca aattcattga ctaaatcttt attgagggtc
  301 tgttacaggc actttattaa ctaataatca gcataatttc tgtgtgagaa taaatgtaaa
  361 aatctgtatt aaaatttcca aatgattatt ttaaatgtat aatgcatgct ctaacagtat
  421 gcccatgtag agctccagag tttttttcttg gaaacagaat gagtagtaca tgagattttc
  481 tgcctcattg gagtagtatt gaagataatt aatataaagg gaaattgtat atttactgat
  541 taattgatat caatctatta attccaacaa gtgaatgtct ctggaaagat tatcaaggca
  601 aagtgttaaa ttggcaaact aaagtcatcc aaaccttcat ttttctgctc acagtgttga
  661 taattaatca gaaaaaagag caaaaaatat taaggtaatt tgaaacaaag tatgttataa
  721 catactatgt ttttatata tttttatatt agaattgaaa tattcagtat ttcttttaca
  781 aaattttct ttcaaaatgt atactttttt ttcttaattt tttttttgc agcttctcat
  841 ggtcaagaat gtatactatt ctgtgggcta aatatcatat cttagaatta taagacatag
  901 aaacattaaa tgaatagaga taaactcagg tgtaaattat gcaattaaaa tggactgcat
  961 tctattatgc atttaactaa ggtcattttt tttttaatgc acaaaaagaa acacccaaaa
 1021 gatatatctg gaaaactttc tttcagtgat acattttttcc tgtttttttt ctgctttcta
 1081 tttgtttaat ag
//
```

**Figure 7.** An Intron sequence from Humretblas Genome

The shortness of the average length of exons (114 bp) used from Humretblas as compared to that of the introns (2347 bp) is a fundamental limitation for this analysis. To make a reasonable comparison, we have used a total coding sequence by combining all exons.

| Sequence | Length (bp) | H3 | H4 | H5 | H6 | H7 | H8 | H9 |
|---|---|---|---|---|---|---|---|---|
| HUMRETBLAS | 180388 | | | | | | | |
| Exon 1 | 197 | 1.71 | 1.30 | 1.05 | 0.82 | 0.6868 | 0.5731 | 0.4878 |
| Exon 2 | 137 | 1.35 | 1.13 | 0.92 | 0.739 | 0.59 | 0.5 | 0.43 |
| Exon 3 | 127 | 1.49 | 1.22 | 0.93 | 0.73 | 0.60 | 0.49 | 0.42 |
| Exon 4 | 116 | 1.53 | 1.21 | 0.89 | 0.71 | 0.57 | 0.47 | 0.40 |
| Exon 5 | 68 | 1.32 | 1.02 | 0.74 | 0.58 | 0.45 | 0.38 | 0.31 |
| Exon 6 | 39 | 1.13 | 0.79 | 0.56 | 0.43 | 0.33 | 0.25 | 0.25 |
| Intron 1 | 3227 | 1.94 | 1.89 | 1.71 | 1.48 | 1.26 | 1.08 | 0.95 |
| Intron 2 | 2687 | 1.86 | 1.80 | 1.65 | 1.43 | 1.21 | 1.04 | 0.91 |
| Intron 3 | 2622 | 1.85 | 1.79 | 1.63 | 1.41 | 1.20 | 1.04 | 0.90 |
| Intron 4 | 2522 | 1.88 | 1.81 | 1.67 | 1.42 | 1.20 | 1.04 | 0.90 |
| Intron 5 | 1936 | 1.88 | 1.78 | 1.59 | 1.35 | 1.15 | 0.98 | 0.86 |
| Intron 6 | 1092 | 1.77 | 1.67 | 1.45 | 1.23 | 1.03 | 0.88 | 0.77 |
| Total coding region | 2787 | 1.87 | 1.851 | 1.68 | 1.44 | 1.22 | 1.05 | 0.92 |

**Figure 8.** Higher-order Exon and Intron Entropy Estimates for HUMRETBLAS

Length of Exon = 197 bp

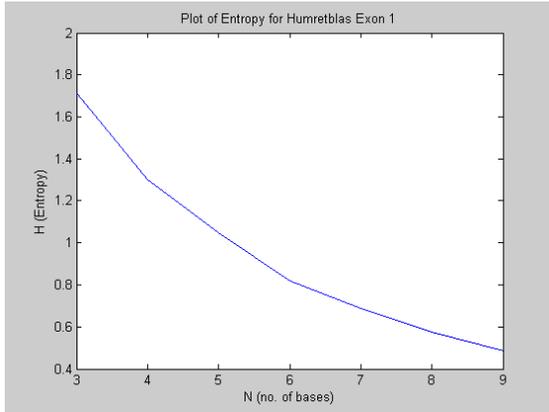

Length of Exon = 137 bp

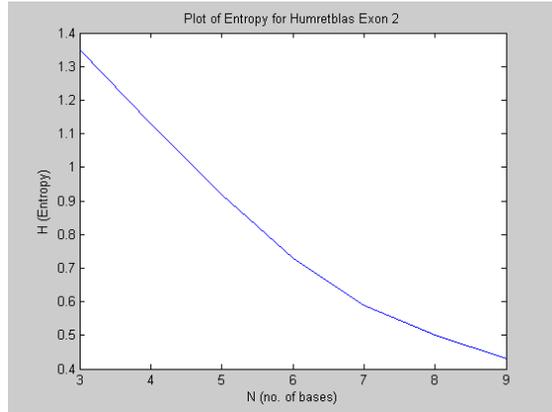

Length of Exon = 127 bp

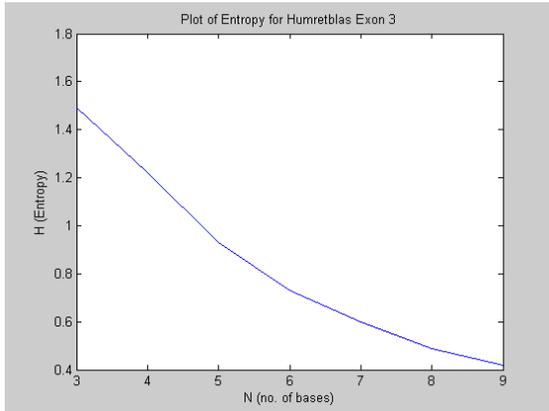

Length of Exon = 116 bp

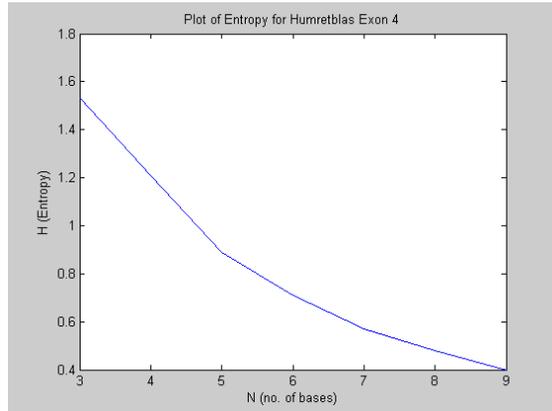

Length of Exon = 68 bp

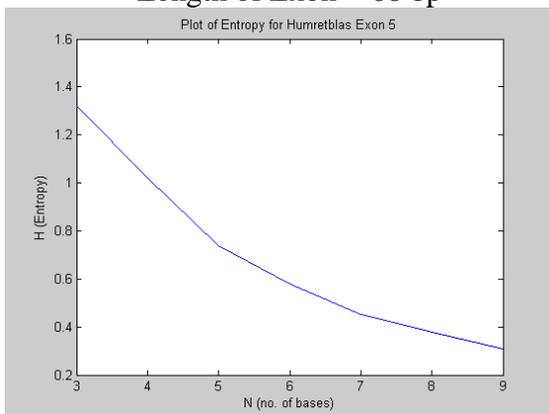

Length of Exon = 39 bp

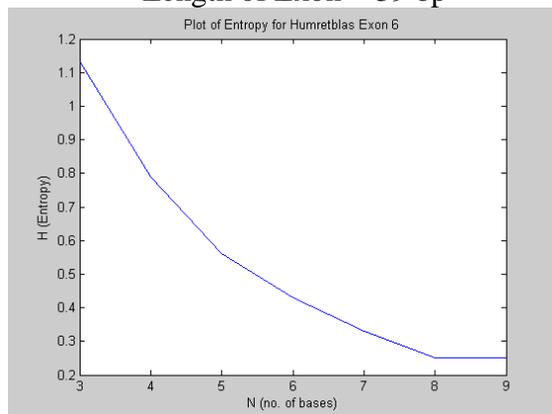

**Figure 9.** Exon Entropy Plots for Humretblas genome

**Entropy Estimate with Nucleotide Pairs**

The smallest sub-component of a DNA sequence was a codon based on its significance from the genetic code. The Universal genetic code indicates that every codon in the DNA sequence maps to a corresponding amino acid. This makes us believe that codons are the primary information carrying strings. There are two key motivations of carrying out this part of the work. The fact that the actual coding sequence is a combination of several exons (DNA Translation) makes it interesting to look at sub-strings other than the codons. Another motivation was derived from an observation of the mapping in the genetic code. As we know, each of the 64 codons map to one of the 20 amino acids that comprise a protein sequence. The Universal genetic code of an organism is presented again along with the observation of codon positions as below.

CT<x> = Leu S

TC<x> = Ser S

| | | T | C | A | G |
|---|---|---|---|---|---|
| T | | TTT Phe (F) | TCT Ser (S) | TAT Tyr (Y) | TGT Cys (C) |
| | | TTC   ” | TCC   ” | TAC | TGC |
| | | TTA Leu (L) | TCA   ” | TAA Ter | TGA Ter |
| | | TTG   ” | TCG   ” | TAG Ter | TGG Trp (W) |
| C | | CTT Leu (L) | CCT Pro (P) | CGT Arg (R) | CGT Arg (R) |
| | | CTC   ” | CCC ” | CGC ” | CGC ” |
| | | CTA   ” | CCA ” | CGA ” | CGA ” |
| | | CTG   ” | CCG ” | CGG ” | CGG ” |
| A | | ATT Ile (I) | ACT Thr (T) | AAA Asn (N) | AGT Ser (S) |
| | | ATC ” | ACC ” | AAC ” | AGC ” |
| | | ATA ” | ACA ” | AAA Lys (K) | AGA Arg (R) |
| | | ATG Met (M) | ACG ” | AAG ” | AGG ” |
| G | | GTT Val (V) | GCT Ala (A) | GAT Asp (D) | GGT Gly (G) |
| | | GTC ” | GCC ” | GAC ” | GGC ” |
| | | GTA ” | GCA ” | GAA Glu (E) | GGA ” |
| | | GTG ” | GCG ” | GAG ” | GGG ” |

**Figure 10.** Behavior of nucleotide pairs from Genetic Code

As see from the genetic code table, pairs of nucleotides are sometimes sufficient to encode for an amino acid which makes the third codon position seem redundant. For example, the codon CC<x> codes for Pro (P) irrespective of the base value that <x> assumes. Although there are exceptions to this in other parts of the genetic code, it holds for a majority of pairs. Intrigued by this behavior of pairs of nucleotides, it might be appropriate to also look at the DNA sequence as a sequence of pairs of nucleotides. There are $4^2 = 16$ possible pairs of nucleotides for our alphabet {A,T,G,C} and the entropy of pairs was normalized by a factor 2 i.e., $H(G) = H(G)'/2$ to obtain the entropy per codon and hence achieving entropy plots with the first entropies obtained using pairs.

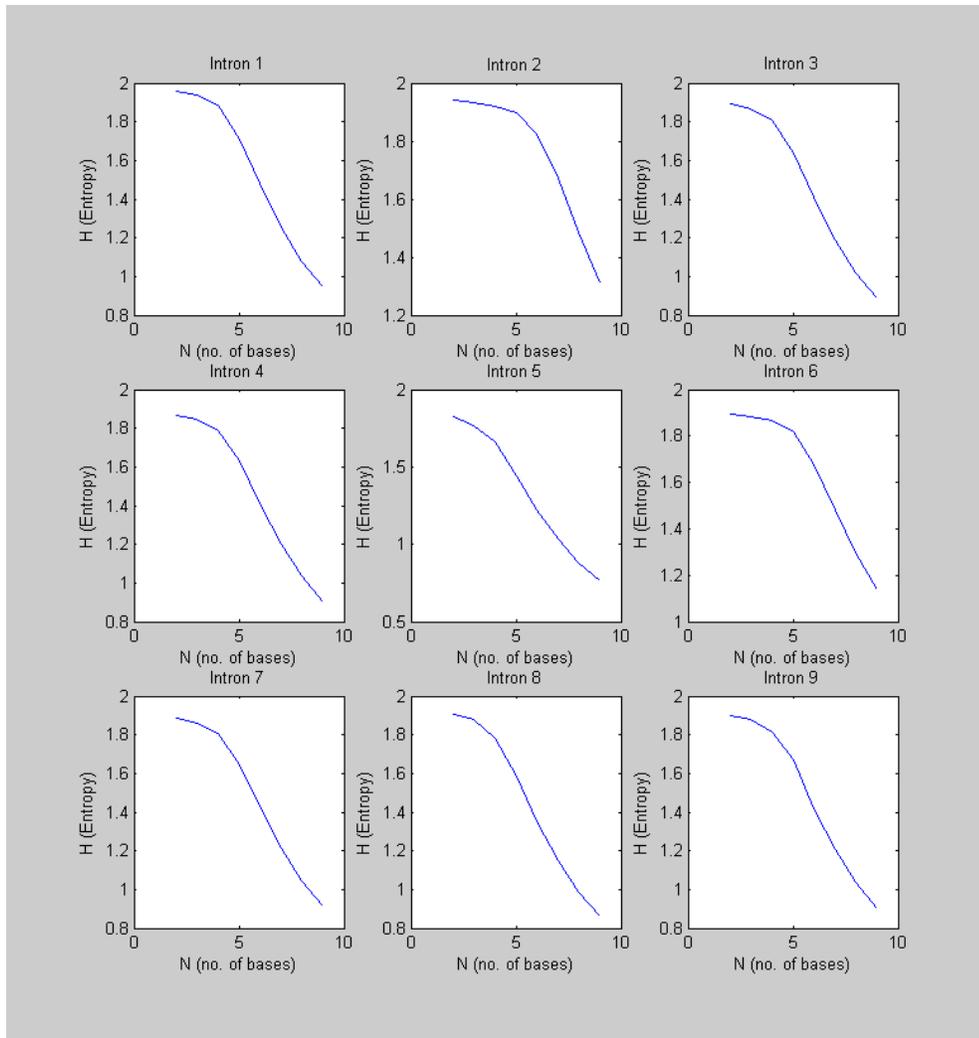

**Figure 11**. Intron entropy plots for Humretblas including entropies estimated using pairs

The entropy values calculated using a sub-string length of 2 are almost equal to the entropy values found using triplets for the sequences tested. The entropy variation plots above include the entropy estimate using base pairs. The entropy due to pairs doesn't seem to affect the rate of entropy convergence seems to be the same for all sequences used. From one perspective, this perhaps further emphasizes a fast rate entropy convergence as a universal property of all intron and exon sequences using our estimator. The finiteness of the exons might have a significance effect on the entropy. This also calls for a technique to take the effect of finiteness into account and hence support our exploration of entropy variation.

### Benchmarking of Entropy Using an Ensemble of Random Sequences

As the length of the sub-string L used in entropy calculation increases so does the set of possible L-tuples. A larger sequence of length exceeding the number of L-tuples is more likely give an accurate estimate. However, due to shortness of the exon sequences, a technique is needed to make a reasonable study of their entropy. Previously attempts have been made to derive the onset of finite sample effects on entropy estimation under the assumption that rank ordered distributions tend to follow Zipf's law [18].

In order to address the problem of finiteness of the sequence, a proportionate change is made to entropy values of the sequence under study.

The same entropy estimation method was applied to random character sequences equal in length to each of the sequences observed and comprised of uniformly distributed characters from the alphabet $\{A, T, G, C\}$. By using simple MATLAB codes, random sequences from the character $\{A, T, G, C\}$ and length equal to the DNA character sequence under question are generated. The entropy values for an ensemble of random sequences of the same length show entropy convergence. This may well be attributed to the finiteness of the sequences.

Since the random sequences are also from the same alphabet, we can use the same entropy algorithm get the entropy values $H(R_i)$ using randomly generate sequences from the alphabet $\{A, T, G, C\}$. From the above results, entropy values decrease with increase in search string length one can attribute this in part to the finiteness of the character sequence. The proportion of the random sequence is calculated as,

$$\Delta_i = \frac{2L}{H(R_i)} \tag{17}$$

Using an ensemble of such random sequences, we calculate the average $H(R_i)$. The correction in the entropy value of sequence under test is done by a multiplying the entropy $H(G_i)$ with the corresponding proportion $\Delta_i$ to get the corrected entropy $\hat{H}(G_i)$ i.e.,

$$\hat{H}(G_i) = H(G_i) * \Delta_i \tag{18}$$

The benchmarked entropies for intron and exon sequences of the Humretblas genome are as illustrated in the table below.

| Sequence | Length (bp) | H3 | H4 | H5 | H6 | H7 | H8 | H9 |
|----------|-------------|------|------|------|------|------|------|------|
| HUMRETBLAS | 180388 | | | | | | | |
| Exon 1 | 197 | 1.98 | 1.93 | 1.99 | 1.97 | 2 | 2 | 1.97 |
| Exon 2 | 137 | 1.67 | 1.82 | 1.94 | 1.93 | 1.95 | 1.94 | 2 |
| Exon 3 | 127 | 1.84 | 2.03 | 2.02 | 2 | 2 | 1.95 | 2 |
| Exon 4 | 116 | 1.91 | 2.04 | 1.98 | 2 | 2 | 2 | 2 |
| Exon 5 | 68 | 1.92 | 2 | 2 | 2 | 2 | 2 | 2 |
| Exon 6 | 39 | 1.91 | 2 | 2 | 2 | 2 | 2 | 2 |
| Intron 1 | 3227 | 1.95 | 1.94 | 1.95 | 1.99 | 1.94 | 1.99 | 1.93 |
| Intron 2 | 2687 | 1.88 | 1.88 | 1.92 | 1.98 | 1.98 | 1.99 | 2 |
| Intron 3 | 2622 | 1.81 | 1.84 | 1.92 | 1.97 | 1.98 | 1.98 | 1.99 |
| Intron 4 | 2522 | 1.90 | 1.89 | 1.97 | 1.99 | 2.1 | 2.01 | 2 |
| Intron 5 | 1936 | 1.90 | 1.88 | 1.92 | 1.96 | 1.99 | 1.99 | 2 |
| Intron 6 | 1092 | 1.82 | 1.85 | 1.92 | 1.97 | 1.98 | 1.99 | 2 |
| Total coding region | 2787 | 1.89 | 1.92 | 1.94 | 1.98 | 1.98 | 2 | 2 |

**Figure 12.** Benchmarked Entropy Results for Humretblas

Entropy of majority of sequences that were used here show that the entropy at L = 4 is either a peak or equal to the codon entropy and followed by a steady fall in the slope. This implies that a string of 4 characters in a DNA character sequence carries an entropy value higher than that of triplets and all strings of a higher length. It makes an interesting observation that strings of length four carry greater information in comparison to codons that play a major role in the transcription of mRNA to protein sequences as seen earlier. This result is perhaps a consequence of redundancy in certain codon positions. The role of the genetic code has been under study and it is believed that there may be more unknown information present within DNA sequences that plays a role in the conversion of DNA to proteins.

## SIGNIFICANCE OF THE WORK

### About Intron Sequences

Considering the fact that introns are finite character sequences, we have an opportunity of exploring their structure, statistical behavior and patterns in comparison to other sequences. In this chapter, we will present the results in our attempt to study the behavior of intronic sequences. We would like to recall the earlier argument that it is necessary to account for the length of the DNA sequences, in addition, to their complexity, structure and sequence pattern, in order to apply information theoretic concepts for their analysis. It is a well known observation that exons tend to be around 200 characters long while introns can stretch to as many as tens of thousands of characters in length [9]. Although the observation may only be an approximation, it is in conformity with the well known fact that only about 3% of the entire genome actually codes for proteins and the remaining is introns and non-coding DNA. Just as the codons indicate the beginning and end of exons, the introns have mostly occurring start and stop indicators as being GT and AG respectively [9]. However many other locations can resemble such patterns and hence these indicators cannot solely suffice to recognize a splice junction. Researchers have attempted using information theoretic techniques like entropy to differentiate between the two finite character sequences – exons and introns. It has been said that entropy is a useful tool in the analysis of DNA sequences [9]. In order to present a rational point of view on the characteristic of such complex, high capacity information storage sequences as DNA sequences we would need a flawless framework.

### Intronic Entropy Results

With the intent of analyzing the effectiveness of entropy as a measure, let us take a quick look back at how entropy of the sequences was calculated here. The entropy calculation is based on Shannon's definition,

$$H(X) = -\sum_i p(x_i) \log p(x_i) \qquad (19)$$

This can be represented for DNA sequences in terms of the constituent strings $g_i$ as,

$$H(S) = -\sum_{i-1}^{N/L} p(g_i) \log p(g_i) \qquad (20)$$

where $S$ is the DNA sequence under question which in this case will be either introns or exons and $g_i$ is a search sequence formed of characters from $\{A, T, G, C\}$ of length $L$ and $N$ is the length of the DNA sequence.

Length of Intron = 3227                Length of Intron = 2687

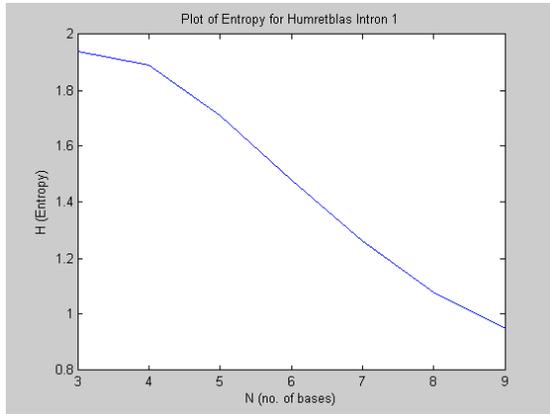
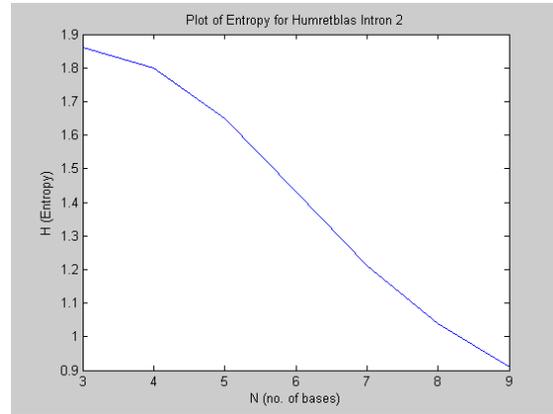

Length of Intron = 2622                Length of Intron = 2522

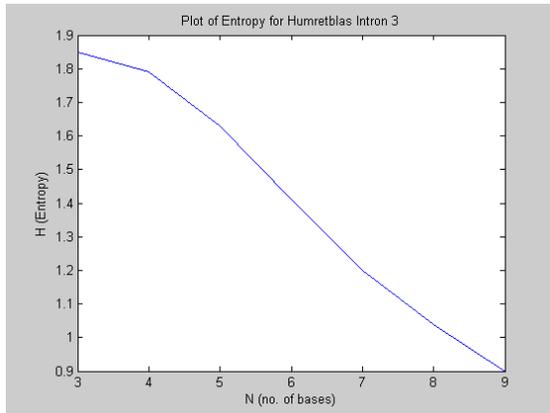
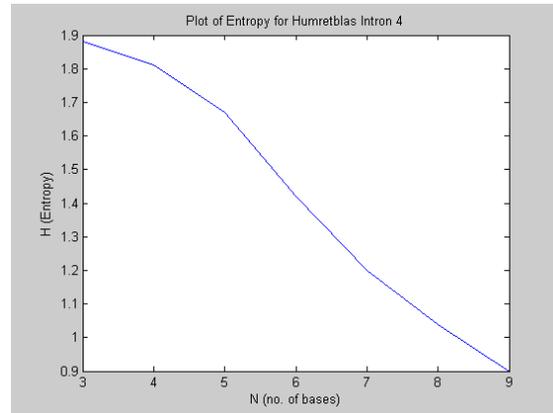

**Figure 13.** Entropy plots of Intron sequences

Length of Intron = 1936

Length of Intron = 1092

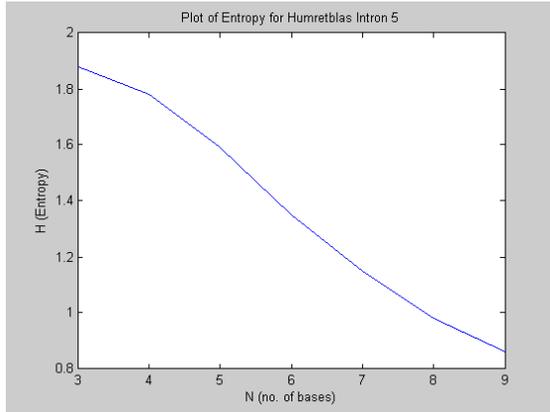
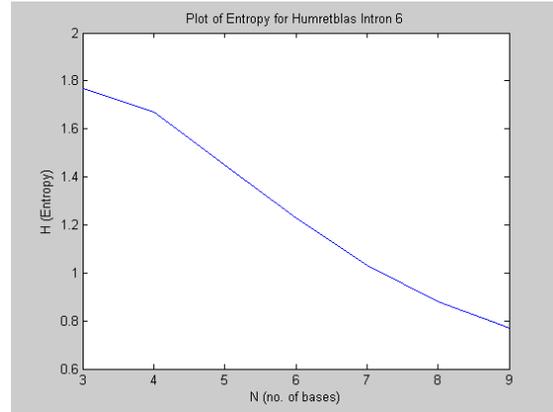

Length of sequence = 10986 bases

Length of sequence = 33895 bases

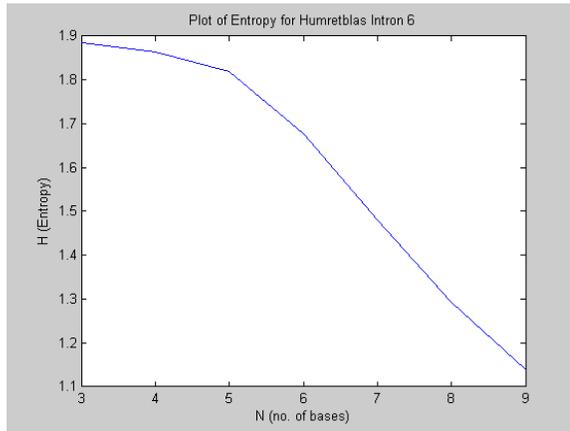
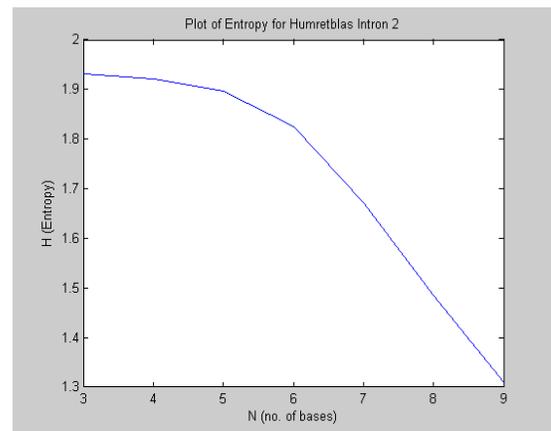

**Figure 13.** Entropy plots of Intron sequences (contd.)

The pattern of entropy convergence in intron sequences seems to match that of the exons in spite of the much higher lengths of introns than the exons. The limited amount of exons is expected due to the presence of only 5% of coding DNA in an organism. Introns have high entropy values as close to 1.97 and 1.98 using codons and in some cases are seen to fall below 1. This might indicate they have some kind of underlying structure. It has been observed earlier that entropy values of the total coding region are very comparable to the intron sequence of equivalent length. This result is likely to be useful in determining similarity between intron and exon sequences.

**CONCLUSION**

The entropy of a number of DNA coding and non-coding sequences collected from different genomes was estimated using a frequency based entropy estimation algorithm for finite sequences. The exon and intron entropy plots both converge in value with increase in length in a similar fashion. For bench-marking, the same entropy estimation method was applied to random character sequences equal in length to each of the sequences tested; the bench-marking sequence comprised of uniformly distributed characters from the alphabet $\{A, T, G, C\}$. In order to deal with the problem of finiteness of the sequence and to make a reasonable entropy comparison between intron and exon sequences that come in different lengths, a correction factor was obtained for every exon/intron sequence using an ensemble of random sequences of the same length. Entropy plots of some of the sequences show a peak at L = 4 followed by a steady fall in the slope. This implies that the bases in a string of 4 characters in a DNA sequence carry average information higher than that for triplets and all strings of a higher length. This is one of the significant findings of this thesis, indicating that least correlation occurs across adjacent codons and that there exist stronger correlation beyond. The relationship across codons was captured most clearly when the normalization using the benchmark sequence results was done, and it was found that both exons and introns have such long-range correlations. This finding is likely to be useful in further understanding of the nature of the genetic code.

The similarity in the entropy of exons and introns suggests that the introns are likely to be playing some hitherto unknown but useful role. Since entropy is directly related to the information content, the similarity of entropy patterns indicates that introns have hidden information. It is not known if this information is useful in repair of exons, or for some independent function. We hope that future work would seek correlations of this information with cellular function.